\documentclass[acus]{JAC2003}
\usepackage{url}

\usepackage{graphicx}

\makeatletter
\newcommand\ps@repno{%
  \renewcommand\@oddhead{\hbox to\textwidth{\hfil BNL-78088-2007-CP}}%
  \renewcommand\@evenhead{\@oddhead}%
  \renewcommand\@oddfoot{}%
  \renewcommand\@evenfoot{\@oddfoot}%
}
\makeatother
\begin{document}
\title{A COMPLETE SCHEME OF IONIZATION COOLING FOR A MUON COLLIDER       \thanks{Work supported by US Department of Energy under contracts AC02-98CH10886 and DE-AC02-76CH03000}}

\author{Robert B. Palmer, J. Scott Berg, Richard C. Fernow, Juan Carlos Gallardo, Harold G. Kirk\\
 (BNL, Upton, NY); Yuri Alexahin,  David Neuffer  (Fermilab, Batavia, IL);
 Stephen Alan Kahn\\ (Muons Inc, Batavia, IL);
 Don Summers (University of Mississippi, Oxford, MS)}
\maketitle
\thispagestyle{repno}
\vskip.5in
\begin{abstract}

A complete scheme for production and cooling a muon beam for three specified muon colliders is presented. Parameters for these muon colliders are given. The scheme starts with the front end of a proposed neutrino factory that yields bunch trains of both muon signs. Emittance exchange cooling in slow helical lattices reduces the longitudinal emittance until it becomes possible to merge the trains into single bunches, one of each sign. Further cooling in all dimensions is applied to the single bunches in further slow helical lattices. Final transverse cooling to the required parameters is achieved in 50 T solenoids using high $T_C$ superconductor at 4~K.
Preliminary simulations of each element are presented.
\end{abstract}

\begin{table}[hbt]
\begin{center}
\caption{Parameters of three muon colliders using the proposed cooling scenario.}
\begin{tabular}{|l|ccc|c|}
\hline
  $E_{\rm c~of~m}  $& 1.5&  4& 8&   TeV\\
 $\cal{L}$ &  1 &  4  & 8&    $10^{34}$ cm$^2$sec$^{-1}$\\
\hline
 $\Delta\nu$ & 0.1& 0.1&0.1& \\
$\mu$/bunch& 2& 2& 2& $10^{12}$\\
 $<B_{\rm ring}>$& 5.2& 5.2&10.4&T\\
$\beta^*  = \sigma_z$& 10&3&3&mm\\
rms $dp/p$& 0.09&0.12&0.06&\%\\

  $N_\mu/N_{\mu o}$& 0.07 & 0.07& 0.07&\\
\hline
 Rep.& 13  & 6&3&Hz\\
 $P_{\rm driver}$& $\approx$4 & $\approx$ 1.8& $\approx$ 0.8&MW\\

 $\epsilon_\perp$& 25 &  25& 25& pi mm mrad\\
 $\epsilon_\parallel$ & 72    & 72& 72& pi mm rad\\
\hline
\end{tabular}
\label{parameters}
\end{center}
\end{table}

\begin{figure}[htb]
\centering
\includegraphics*[width=75mm]{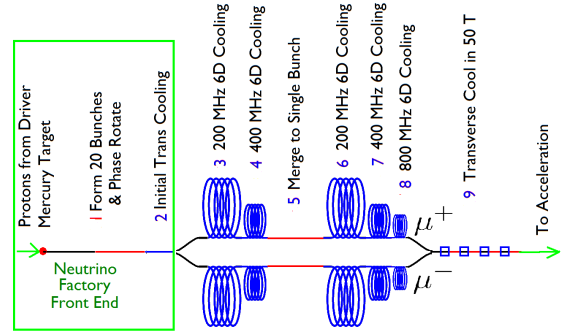}
\caption{Schematic of the components of the muon manipulations and cooling.}
\label{schematic}
\end{figure}

\begin{figure}[htb]
\centering
\includegraphics*[width=75mm]{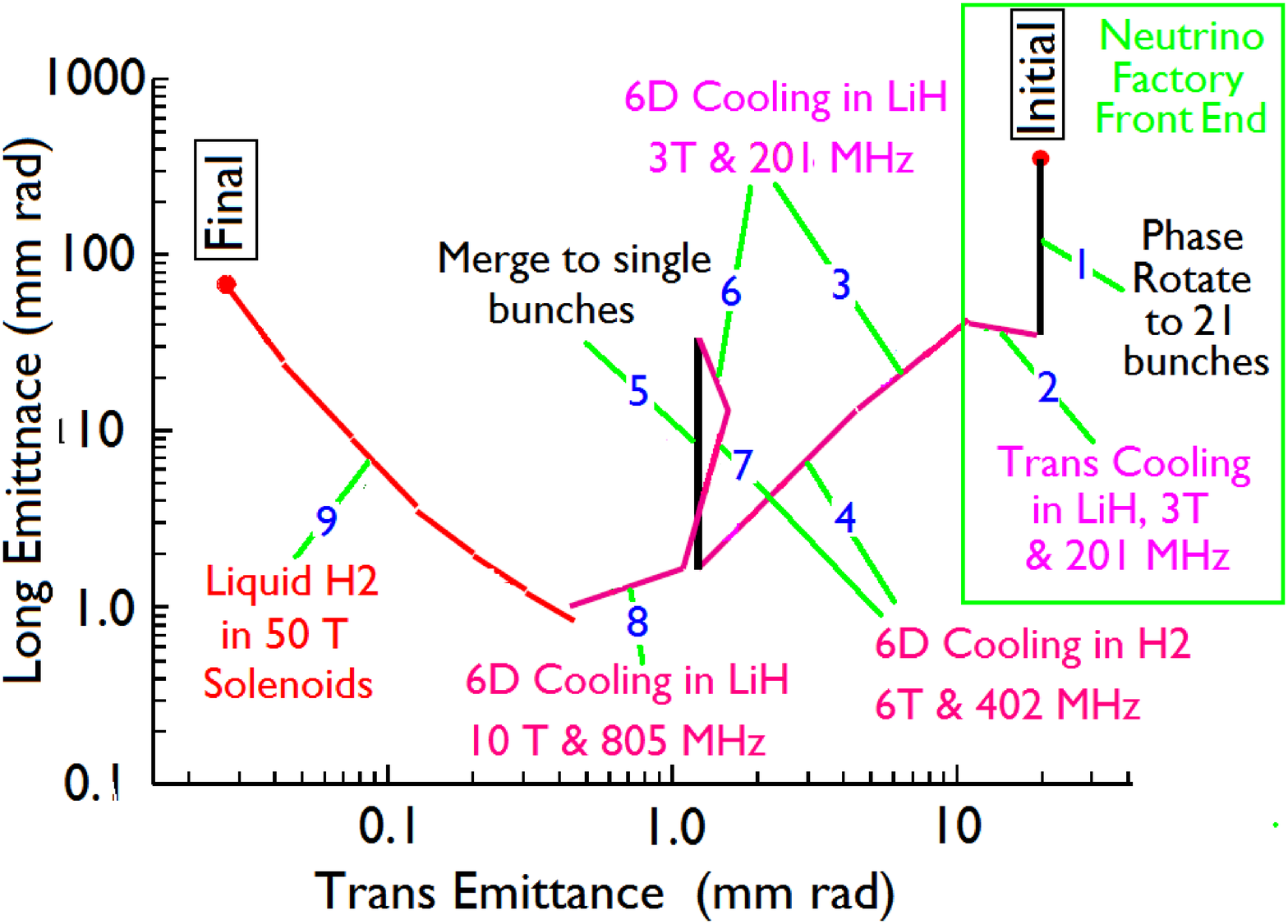}
\caption{Transverse vs. longitudinal emittances before and after each stage}
\label{longtrans}
\end{figure}
\section{INTRODUCTION}

Muon colliders were first proposed by Budker in 1969~\cite{budker}, and later discussed by others~\cite{skrinsky}. A more detailed study was done for Snowmass 96~\cite{feasibility}, but in none of these was a complete scheme defined for the manipulation and cooling of the required muons.  

Muon colliders would allow the high energy study of point-like collisions of leptons without some of the difficulties associated with high energy electrons, such as the synchrotron radiation requiring their acceleration to be essentially linear, and as a result, long.  Muons can be accelerated in smaller rings and offer other advantages, but they are produced only diffusely and they decay rapidly, making the detailed design of such machines difficult.  In this paper, we outline a complete scheme for capture, phase manipulation and cooling of the muons, every component of which has been simulated at some level.

\section{COLLIDER PARAMETERS}

Table ~\ref{parameters} gives parameters for muon colliders at three energies.  Those at 1.5 TeV  correspond to a recent collider ring design~\cite{alexahin}. The 4 TeV example is taken from the 96 study~\cite{feasibility}.  The 8 TeV is an extrapolation assuming higher  bending fields and more challenging interaction point parameters. All three use the same muon intensities and emittances, although the repetition rates for the higher energy machines are reduced to control neutrino radiation.

\section{PROPOSED SYSTEM}
Figure~\ref{schematic} shows a schematic of the components of the  system. Figure~\ref{longtrans} shows a plot of the longitudinal and transverse emittances of the muons as they progress from production to the specified requirements for the colliders.  The subsystems used to manipulate and cool the beams to meet these requirements are indicated by the numerals 1--9 on the figures.


The muons are generated by the decay of pions produced by proton bunches interacting in a mercury jet target.  These pions are captured by a 20 T solenoid surrounding the target, followed by an adiabatic lowering of the field to a decay channel.  

The first manipulation (\#1), referred to as phase rotation~\cite{neuffer-rot}, converts the initial single short muon bunch with very large energy spread into a train of 21 bunches with much reduced energy spread.  The initial bunch is allowed to lengthen and develop a time-energy correlation in a 110~m drift.  It is then bunched into a train, without changing the time-energy correlation, using rf cavities whose frequency varies with location, falling from 333 MHz to 234~MHz.  Then, by phase and frequency control, the rf accelerates the low energy bunches and decelerates the high energy ones.  Muons of both signs are captured and then (\#2) cooled transversely in a linear channel using LiH absorbers, periodic alternating 2.8 T solenoids, and 201 MHz rf.  All the components up to this point are identical to those described in a recent study~\cite{study2a} for a neutrino factory.

\begin{figure}[htb]
\centering
\includegraphics*[width=70mm]{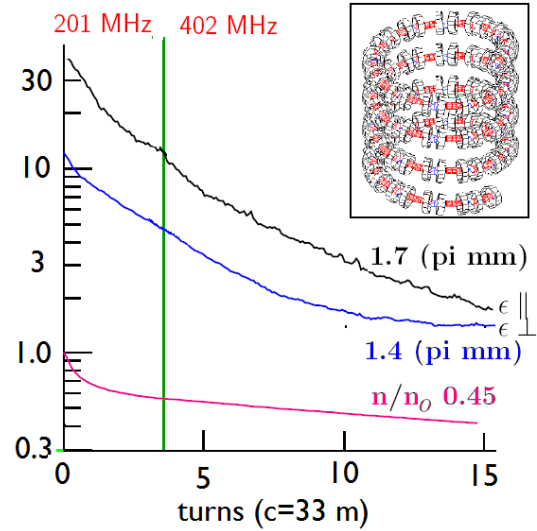}
\caption{ICOOL simulation of 6D cooling in stages \#3 \& \#4. Inset: Geometry of \#3}
\label{rfofo34}
\end{figure}

The next stage (\#3) cools simultaneously in all 6 dimensions.  The lattice~\cite{rfofo} uses 3 T solenoids for focus, weak dipoles (generated by tilting the solenoids) to generate dispersion, wedge shaped liquid hydrogen filled absorbers where the cooling takes place, and 201 MHz rf, to replenish the energy lost in the absorbers.  The dipole fields cause the lattices to curve, forming a slow upward or downward helix (see inset in fig.~\ref{rfofo34}).
The following stage (\#4) uses a lattice essentially the same as \#3, but with twice the field strength, half the geometric dimensions, and 402 instead of 201 MHz rf.  Fig.~\ref{rfofo34} shows the results of a simulation of both systems using ICOOL.  Although this simulation was done for circular, rather than the helical, geometry, it used realistic coil and rf geometries.  Preliminary studies~\cite{guggenheim} suggest that the differences introduced by the helical, instead of circular, geometries will be negligible. The simulation did not include the required matching between the two stages. The simulations also used fields that, while they satisfied Maxwell's equations and had realistic strengths, were not actually calculated from specified coils. Simulations reported in reference~\cite{rfofo}, using fields from actual coils, gave essentially identical results.

\begin{figure}[htb]
\centering
\includegraphics*[width=\linewidth]{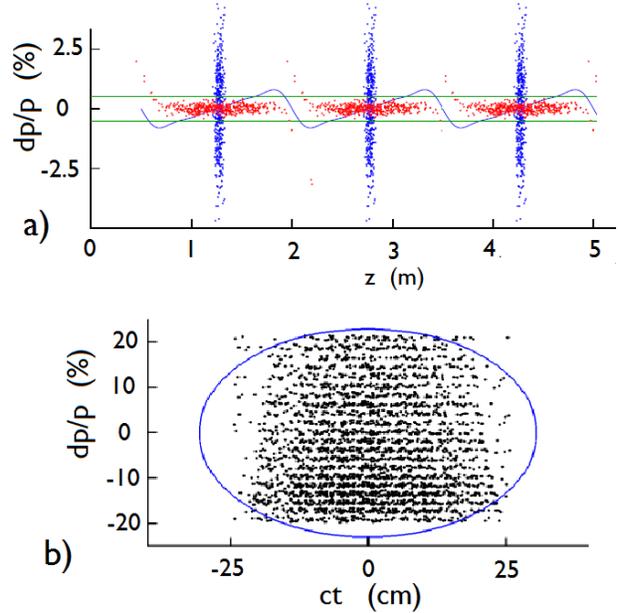}
\caption{1D Simulation of merge (\#5): a) before (blue) and after (red) first rotation, b) after second rotation}
\label{merge}
\end{figure}

Since collider luminosity is proportional to the square of the number of muons per bunch, it is important to use relatively few bunches with many muons per bunch. However, capturing the initial muon phase space into single bunches requires low frequency ($\approx$ 30 MHz) rf, and thus low gradients, resulting in slow initial cooling. It is thus advantageous to capture initially into multiple bunches at 201 MHz and merge them after cooling allows them to be recombined into a single bunch at that frequency.  This recombination (\#5) is done in two stages:  a) using a drift followed by 201 MHz rf, with harmonics, the individual bunches are phase rotated to fill the spaces between bunches and lower their energy spread; followed by b) 5 MHz rf, plus harmonics, interspersed along a long drift to phase rotate the train into a single bunch that can be captured using 201 MHz.  Results of an initial one dimensional simulation of this process is shown in Fig.~\ref{merge}.  Work is ongoing on the design and simulation of a system with the low frequency rf separated from a following drift in a wiggler system with greater momentum compaction to reduce the length and decay losses.

\begin{figure}[tb]
\centering
\includegraphics*[width=75mm]{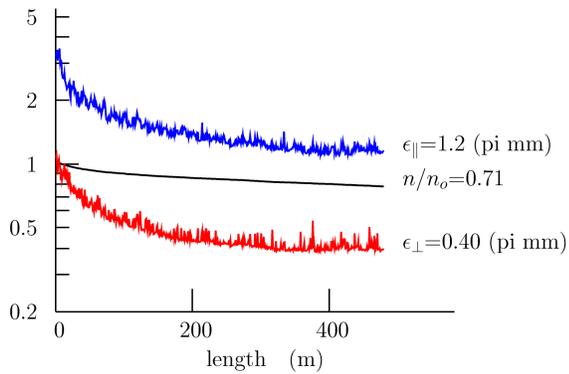}
\caption{ICOOL simulation of final 6D cooling lattice (\#8)
 using 10T solenoids and 805 MHz rf.}
\label{last6d}
\end{figure}

After the bunch merging, the longitudinal emittance of the single bunch is now similar to that at the start of cooling.  It can thus be taken through the same, or similar, cooling systems as \#3 and \#4:  now numbered \#6 and \#7.  One more (\#8) stage of 6 dimensional cooling has been designed, using 10 T magnets, hydrogen wedge absorbers, and 805 MHz rf. Its ICOOL simulated performance is show in Fig.~\ref{last6d}.  Again, the simulation shown used fields that, while they satisfied Maxwell's equations and had realistic strengths, were not actually calculated from specified coils.

\begin{figure}[tb]
\centering
\includegraphics*[width=65mm]{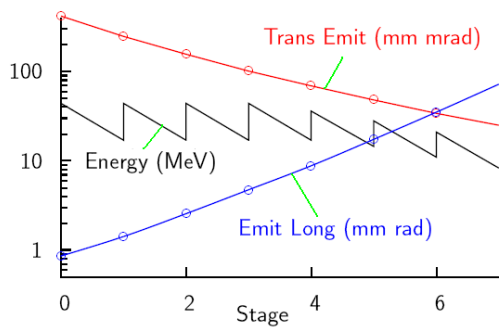}
\caption{Results of ICOOL simulations of transverse cooling in liquid hydrogen in 7 sequential 50 T solenoids}
\label{50T}
\end{figure}



To attain the required final  transverse emittance, the cooling needs stronger focusing than is achievable in the 6D cooling lattices used in the earlier stages. It can be obtained in liquid hydrogen in strong solenoids, if the momentum is allowed to fall, but at the lower momenta the momentum spread, and thus longitudinal emittance, rises relatively rapidly. However, as we see from 
Fig.~\ref{longtrans}, the longitudinal emittance after \#8 is far less than that required, so such a rise is acceptable. Figure~\ref{50T} shows the results of ICOOL simulation of cooling in seven 50 T solenoids. The simulation did not include the required matching and re-accelerations between the solenoids.

\section{ONGOING STUDIES}

The calculated space charge tune shifts are moderate, but space charge is not yet in the simulations. There is a question as to whether the specified gradients of rf cavities operating under vacuum would operate in the specified magnetic fields. This is under study by our collaboration and alternative designs using high pressure hydrogen gas, or open cell rf with solenoids in the irises, are being considered.  The bunching and phase rotation should be optimized for the muon collider, instead of being copied from a neutrino factory.  Instead of the slow helices, a planar wiggler lattice is being studied that would cool both muon signs simultaneously, thus greatly simplifying the system. The use of more, but lower field (e.g., 35 T) final cooling solenoids is also under study. In addition, many details need designing and simulating, and the various new technologies (mercury target, ionization cooling, helical lattice, high field solenoids, etc.) need demonstrating.

\section{CONCLUSION}

Although much work remains to be done, the scenario outlined here appears to be a plausible solution to the problems of capturing, manipulating, and cooling muons to the specifications for muon colliders with useful luminosities and energies, even up to 8 TeV in the center of mass.

\end{document}